\documentclass[a4paper]{jpconf}
\usepackage{graphicx}
\begin{document}
\title{Phase diagram of neutron-rich nuclear matter and its impact on 
astrophysics}

\author{F Gulminelli$^1$, Ad R Raduta$^2$, J Margueron$^3$, 
P Papakonstantinou$^3$ and M Oertel$^4$ }

\address{$^1$ CNRS, ENSICAEN, UMR6534, LPC ,F-14050 Caen c\'edex, France}
\address{$^2$ NIPNE, Bucharest-Magurele, POB-MG6, Romania}
\address{$^3$ IPNO, CNRS/IN2P3, Universit\'e Paris - Sud 11, F-91406 
Orsay cedex, France}
\address{$^4$ LUTH, CNRS, Observatoire de Paris, 
Universit\'e Paris Diderot, 5 place
Jules Janssen, 92195 Meudon, France}
   
\ead{gulminelli@lpccaen.in2p3.fr}

\begin{abstract}
Dense matter as it can be found in core-collapse supernovae and
neutron stars is expected to exhibit different phase transitions which
impact the matter composition and equation of state, with important
consequences on the dynamics of core-collapse supernova explosion and
on the structure of neutron stars.  In this paper we will address the
specific phenomenology of two of such transitions, namely the
crust-core solid-liquid transition at sub-saturation density, and the
possible strange transition at super-saturation density in the
presence of hyperonic degrees of freedom.  Concerning the neutron star
crust-core phase transition at zero and finite temperature, it will be
shown that, as a consequence of the presence of long-range Coulomb
interactions, the equivalence of statistical ensembles is violated and
a clusterized phase is expected which is not accessible in the
grand-canonical ensemble.  A specific quasi-particle model will be
introduced to illustrate this anomalous thermodynamics and some
quantitative results relevant for the supernova dynamics will be
shown.  The opening of hyperonic degrees of freedom at higher
densities corresponding to the neutron stars core modifies the
equation of state.
The general characteristics and order of phase
transitions in this regime will be analyzed in the framework of a
self-consistent mean-field approach.
\end{abstract}

\section{Introduction}
Triggered by the first data taking of X and gamma satellites, as well
as by the improved capabilities of radio telescopes and high and
ultra-high gamma ray detectors, there has been in the last ten years
an impressive accumulation of observational data on supernova and
neutron stars in different environments and at different evolution
stages.  Among the most important recent results one should mention
the precise measurement of a very massive two solar-mass neutron
star~\cite{demorest} which challenges a number of theoretical equations
of state of dense matter, and the measurement of the cooling curve of
the young pulsar Cassiopeia A~\cite{cassiopea}, which gives strong
constraints on the super-fluid properties of neutron star cores.

Understanding these data requires a good modelization of the supernova
dynamics and of the formation process of neutron
stars~\cite{nakazato,marek}, but also a precise and detailed control of
the relevant microphysics.  Sophisticated two and three-dimensional
core-collapse supernova simulations show that matter in the supernova
core explores a very large interval of baryonic densities, ranging
from about $\rho>10^{10}$ g $\cdot$ cm$^{-3}\approx 10^{-4}\rho_0$ to
several times the normal nuclear density $\rho_0$ , temperatures
between some hundreds of keV and around 80 
MeV, and proton fractions
between $\approx$ 0.5 and $ \approx$ 0.1.  

 Different microscopic properties of nuclear matter in these wide
 thermodynamic conditions are of influence for these astrophysical
 phenomena.  In particular, the core-collapse gravitational models
 show an important influence of the explosion mechanism with respect
 to the equation of state.  To give an example, the explosion of very
 massive 15 solar-mass progenitors can presently only be achieved with
 a so-called "soft" equation of state. 

In the whole sub-saturation region matter is not
 uniform but it is constituted of finite nuclei, with a dominance of
 exotic neutron-rich isotopes. Therefore a reliable calculation of the
 abundances of these nuclei is essential, as well as a knowledge of
 their mass, level densities and in-medium self-energy modifications.
 
Another key aspect concerns the thermal energy evacuation during the
explosion and the proto-neutron star cooling. In the first evolution
stage, heat is evacuated essentially by electron capture processes
followed by neutrino emission. It is therefore very important to
control the associated electro-weak processes, namely the electron
capture rates and the neutrino interactions with the dense matter and
the different nuclei of the proto-neutron star crust.

Both the equation of state and the neutrino cross sections are strongly 
influenced by the possible presence of phase transitions in dense matter. 
In particular, first-order phase transitions lead to a softening of the 
baryonic pressure, while second order phase transitions lead to a dramatic 
increase of the electroweak interaction probability~\cite{jmargue}.
For this reason, a control of the phase diagram of stellar matter is of 
importance  in our understanding of these complex astrophysical process.

In this contribution, we will limit ourselves to the thermodynamic 
conditions where matter can be described by nucleonic or hadronic degrees 
of freedom and will examine two phase transitions which are expected to play 
a role in supernova physics, namely  the crust-core phase transition at 
sub-saturation density and the possible phase transition towards strange 
matter at super-saturation density.

\section{Sub-saturation densities and the quenching of the first-order phase transition}
Since the early days of the equation of state modelization it was clearly 
recognized that, because of the strong similarities between the effective 
nucleon-nucleon potential and molecular interactions, the phase diagram of 
diluted nuclear matter should present a first-order phase transition of the 
liquid-gas type, terminating at high temperature and density in a critical 
point~\cite{siemens}. This conjecture has been confirmed in all 
phenomenological~\cite{ducoin,rios} or microscopic~\cite{typel} modelizations 
of symmetric and asymmetric nuclear matter with 
realistic effective interactions compatible with recent nuclear data. 

In these calculations, nuclear matter is  defined as an idealized bulk medium 
of neutrons and protons where the electromagnetic interaction is artificially 
switched off in order to achieve a thermodynamic limit.
 In the physical situation of stellar matter, however, charge neutrality is 
achieved by the screening effect of electrons on the proton charge. 
Because of the very different mass between electrons and protons, the 
compressibility of electron and proton matter is very different, which induces 
Coulomb effects that drastically modify the liquid-gas phase transition 
associated to uncharged nuclear matter. Specifically, in all density and 
temperature conditions relevant for neutron star physics, the electron charge 
can be safely considered as uniformly distributed~\cite{maruyama}. 
This means that any baryonic density fluctuation (which is correlated to a 
proton density fluctuation because of the repulsive  symmetry energy) induces 
a fluctuation in the electric charge. A well known consequence of the 
resulting Coulomb correlations is that  the low density phase at zero 
temperature is not a gas, but a Wigner crystal of nuclei immersed in the 
homogeneous electron background~\cite{lattimer}. It is clear that such Coulomb 
effects do not disappear with increasing density and temperature, and it 
is a-priori not  evident that a phenomenology equivalent to the one 
calculated for uncharged nuclear matter might at all be observed. 
However, guided by the uncharged nuclear matter example, 
the standard treatments currently used in most supernovae codes describe 
the dilute stellar matter at finite temperature in the baryonic sector 
as a statistical equilibrium between protons, neutrons, alphas and a single
heavy nucleus representing the finite system counterpart of the liquid 
fraction~\cite{LS91,shen}. The transition to homogeneous matter in the 
neutron star core is supposed to be first-order in these modelizations 
and obtained through a Maxwell construction in the total density at fixed 
proton fraction.

From the nuclear physics side it is well recognized that, since stellar matter 
is subject to the contrasting couplings of the electromagnetic and the strong 
interaction acting on comparable length scales because of  the electron 
screening, this should give rise to the phenomenon of 
frustration~\cite{horowitz}, well-known in 
condensed matter physics~\cite{campa}. Because of this, a specific phase 
diagram, different from the one of nuclear matter and including 
in-homogeneous components, is expected in stellar matter~\cite{noi}. 

\begin{figure}
\begin{center}
\includegraphics[width=0.5\columnwidth,height=4.in,clip]{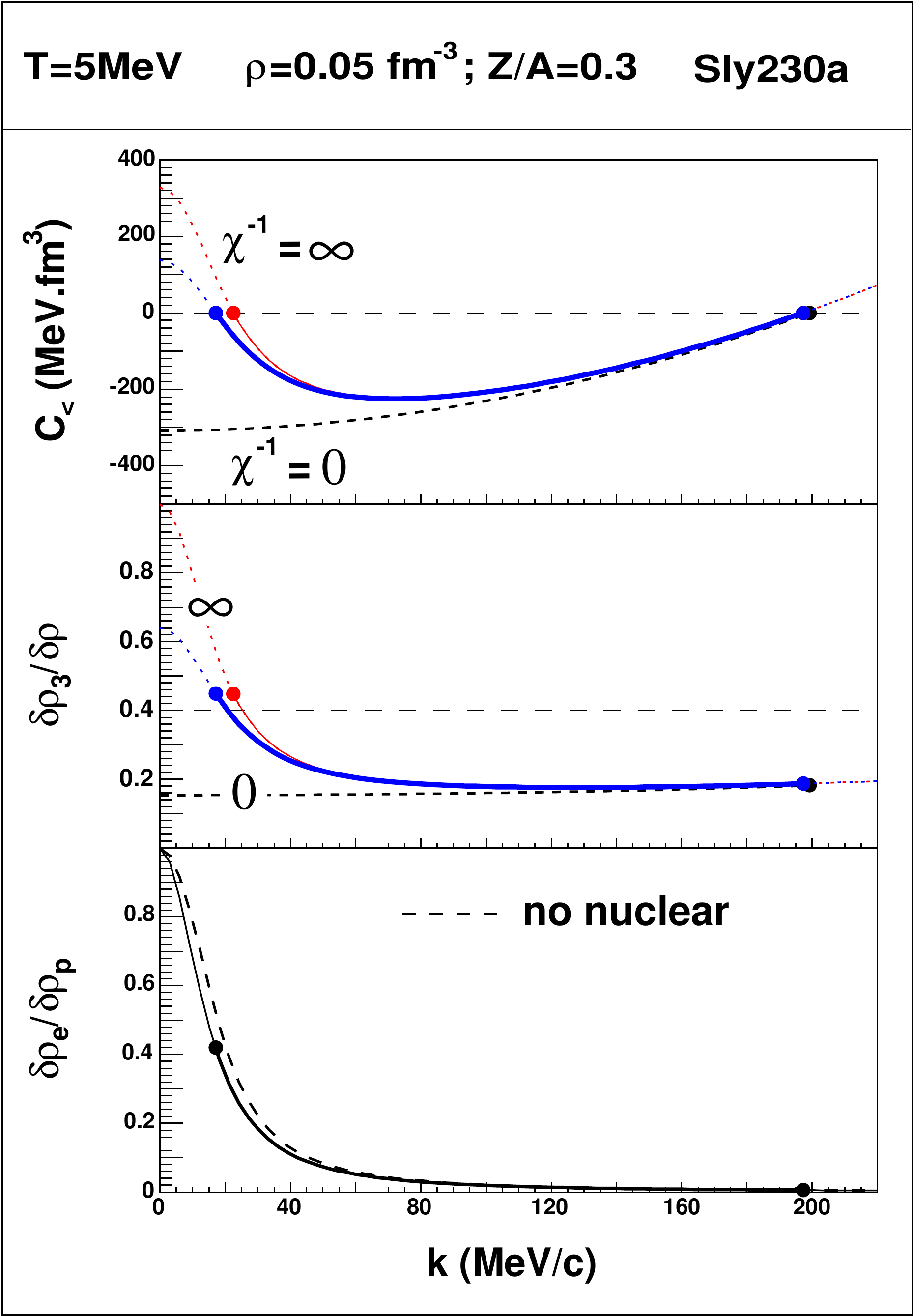}
\caption{ Eigen-mode of the minimal free-energy curvature for 
$\rho$ = 0.05 fm$^{−3}$, Z/A = 0.3, T = 5 MeV, as a function of the 
wave number k, calculated with the Sly230a Skyrme interaction. 
Top: eigen-value. Middle: associated eigen-vector in the nuclear-density 
plane. The curves are compared to two limiting cases corresponding to a 
zero and infinite incompressibility of the negatively charged gas. 
When the curvature is negative (full lines) this eigen-vector gives the 
phase-separation direction. The dots give the points of zero curvature. 
The dashed line gives the direction of constant Z/A. Bottom: 
same as the middle part, in the plane
of proton and electron density. The dashed line gives the eigenvector when 
the nuclear force is zero.  Figure  taken from ref.~\cite{ducoin1}.
 }
\label{fig1}
\end{center}
\end{figure}
\subsection{Coulomb effects in dilute stellar matter}
The quenching of the first-order phase transition due to Coulomb effects 
is illustrated in Figure \ref{fig1} in a finite temperature Hartree-Fock 
calculation with realistic phenomenological interactions~\cite{ducoin1}.
In order to determine the possible first-order phase transition in stellar 
matter, the instability of such matter with respect to a density fluctuation 
can be studied by considering the free energy variation once 
independent proton, neutron, and electron fluctuations $(q,p,e)$
\begin{equation}
\delta \rho_q = A_q e^{i\vec k \cdot \vec r} + c.c.,
\end{equation}
are imposed in a given thermodynamic condition defined by a density, 
temperature, and proton fraction. This variation
 in the three-dimensional space of 
density fluctuations $\tilde{A}=\left ( A_n, A_p, A_e\right )$  
reads~\cite{pethick} : 
\begin{equation}
\delta f={\tilde{A}^*} \mathcal C^f \tilde{A},
\end{equation}
where
%
%
%
\begin{equation}
\label{EQ:Cf}
\mathcal C^f=
\left( 
\begin{array}{ccc}
\partial\mu _{n}/\partial\rho _{n} & \partial\mu _{n}/\partial\rho _{p} & 0\\ 
\partial\mu _{p}/\partial\rho _{n} & \partial\mu _{p}/\partial\rho _{p} & 0\\
0 & 0 & \partial\mu _{e}/\partial\rho _{e}\\ 
\end{array}
\right)
+
k^2
\left( 
\begin{array}{ccc}
2C_{nn}^{f} & 2C_{np}^{f} & 0\\ 
2C_{pn}^{f} & 2C_{pp}^{f} & 0\\
0 & 0 & 0\\ 
\end{array}
\right) 
+
\frac{4\pi^2e^2}{k^2}
\left( 
\begin{array}{ccc}
0 & 0 & 0\\ 
0 & 1 & -1\\
0 & -1 & 1\\ 
\end{array}
\right) 
\end{equation}
%
is the free-energy curvature matrix.
An instability against matter clusterization 
corresponds to a negative free-energy curvature in the space of density 
fluctuations.
It is studied through the $k$-dependent curvature matrix $\mathcal C^f$.
 A negative eigenvalue associated to a given wavelength signals that a 
fluctuation characterized by the associated eigenvector will be spontaneously 
amplified, giving rise to cluster formation if the wavelength is finite, 
or phase separation for infinite wavelengths.  This smallest eigenvalue $C_<$ 
 is shown as a function of the wave number in Figure \ref{fig1}. 
We can see that the eigenvalue is positive at k=0, meaning that the 
phase transition is quenched and replaced by cluster formation in 
stellar matter.

This Coulomb frustration effect has been confirmed by different microscopic 
models~\cite{watanabe} .
A continuous evolution from the Coulomb lattice to a homogeneous nuclear 
fluid, passing through the formation of clusters of different sizes and 
strongly deformed dis-homogeneous structures close to the saturation density 
has been reported, both at zero and at finite temperature. 

These calculations are numerically very heavy and a complete thermodynamic 
characterization of stellar matter under the frustrated phase transition has 
not been done yet. Such a task can however be performed in the 
phenomenological  so-called nuclear statistical equilibrium (NSE) 
approaches,  which treat the bound states of
nucleons as new species of quasi-particles~\cite{NSE,mishustin,blinnikov,souza}.

\subsection{Application in the extended NSE model}
Within NSE, the baryonic component of the stellar matter is regarded as a 
statistical equilibrium of neutrons and
protons, the electric charge of the latter being screened by a homogeneous 
electron background.
As a first approximation, one can consider that the system of interacting 
nucleons is equivalent to a system of non-interacting clusters, with the nuclear 
interaction being completely exhausted by clusterization. This simple model 
can describe only diluted matter at $\rho\ll\rho_0$ as it can be found in 
the outer crust of neutron stars,
while nuclear interaction among nucleons and clusters has to be included for 
applications at higher density,when the average inter-particle distance becomes 
comparable to the range of the
force. 
A simple possibility~\cite{raduta,ensineq} 
is to take into account interactions among 
composite clusters  in the simplified form of a hard-sphere excluded volume, 
and interactions among nucleons  in the self-consistent Hartree-Fock 
approximation with a 
phenomenological realistic energy functional fitted on a large number of 
measured mass and radii. 
In this version of the model, the thermodynamics can be calculated in the 
ensemble $(\beta,\mu_I,\rho)$  defined by inverse temperature $\beta=T^{-1}$, 
isovector chemical potential $\mu_I=\mu_n-\mu_p$ and baryonic density 
$\rho=\rho_n+\rho_p$. The total entropy density is given by
\begin{equation}
\sigma_{\beta,\mu_I}^{can} (\rho)= {\ln {\cal Z}_{\beta,\mu_I}^{a=1}(\rho_f)}   
+ \lim_{V\to\infty} \frac{1}{V}\ln {\cal Z}_{\beta,\mu_I}^{a>1}\left(V \rho_{cl}
 \right) \label{monomer_mix}
\end{equation}
where the density repartition between the clustered $\rho_{cl}$ 
and unbound $\rho_f$ component  $\rho=\rho_{cl}+\rho_f$ is uniquely defined 
by the
condition of having a single chemical potential $\mu$ for both components. 
The unbound nucleons partition sum is calculated in the mean-field 
approximation:
\begin{equation}
 {\ln {\cal Z}_{\beta,\mu_I}^{a=1}(\rho)} = {\ln {\cal Z}_{\beta,\mu,\mu_I}^{0} } -
 \beta \left( \frac{\partial}{\partial{\beta}} \ln {\cal Z}_{\beta,\mu,\mu_I}^{0}
  + V\epsilon^{HF} \right)  -\beta\mu\rho V,
\end{equation}
In this equation $\epsilon^{HF}$ is the Hartree-Fock energy and the 
non-interacting part of the partition sum can be expressed 
as a functional of the  kinetic energy density $\tau _{q}$ 
for neutrons ($q$) or protons ($q=p$) : 

\begin{equation}
\ln {\cal Z}_{\beta,\mu,\mu_I}^{0}  
=\frac{2 \beta V}{3} \sum_{q,p} \tau_{q} \frac{\partial \epsilon^{HF}}
{\partial \tau_{q}}   
, 
\end{equation}
where $\tau_q$ are self-consistently derived as a function of the 
associate chemical potential $\mu_q$ and effective mass $m^*_q$.

The partition sum of clusters  for a total particle number $A=V\rho_{cl}$ 
is calculated via a recursive relation:
\begin{equation}
{\cal Z}_{\beta,\mu_I}(A)= \frac{1}{A}\sum_{a=2}^A a \omega_a {\cal
  Z}_{\beta,\mu_I}(A-a) .
\label{zcan}
\end{equation}
where the weight of the different cluster size is given by:
\begin{equation}
\omega_a=   V_F\sum_{i=-a}^a   \left ( \frac{2\pi a m_0}{\beta h^2}\right )^{3/2}     
\exp\left ( -\beta  f^\beta_{a,i} \right ) \exp (\beta i \mu_I )  . \label{isotopes}
\end{equation}
Here, $ f^\beta_{a,i}$ is a phenomenological 
free energy functional including the screening effect of electrons in the
 Wigner-Seitz approximation, and $V_F=V(1-\rho_{cl}/\rho_0)$ is the 
volume fraction available to the clusters \cite{raduta,ensineq} .

\begin{figure}
\includegraphics[width=0.8\columnwidth,height=5.in,clip]{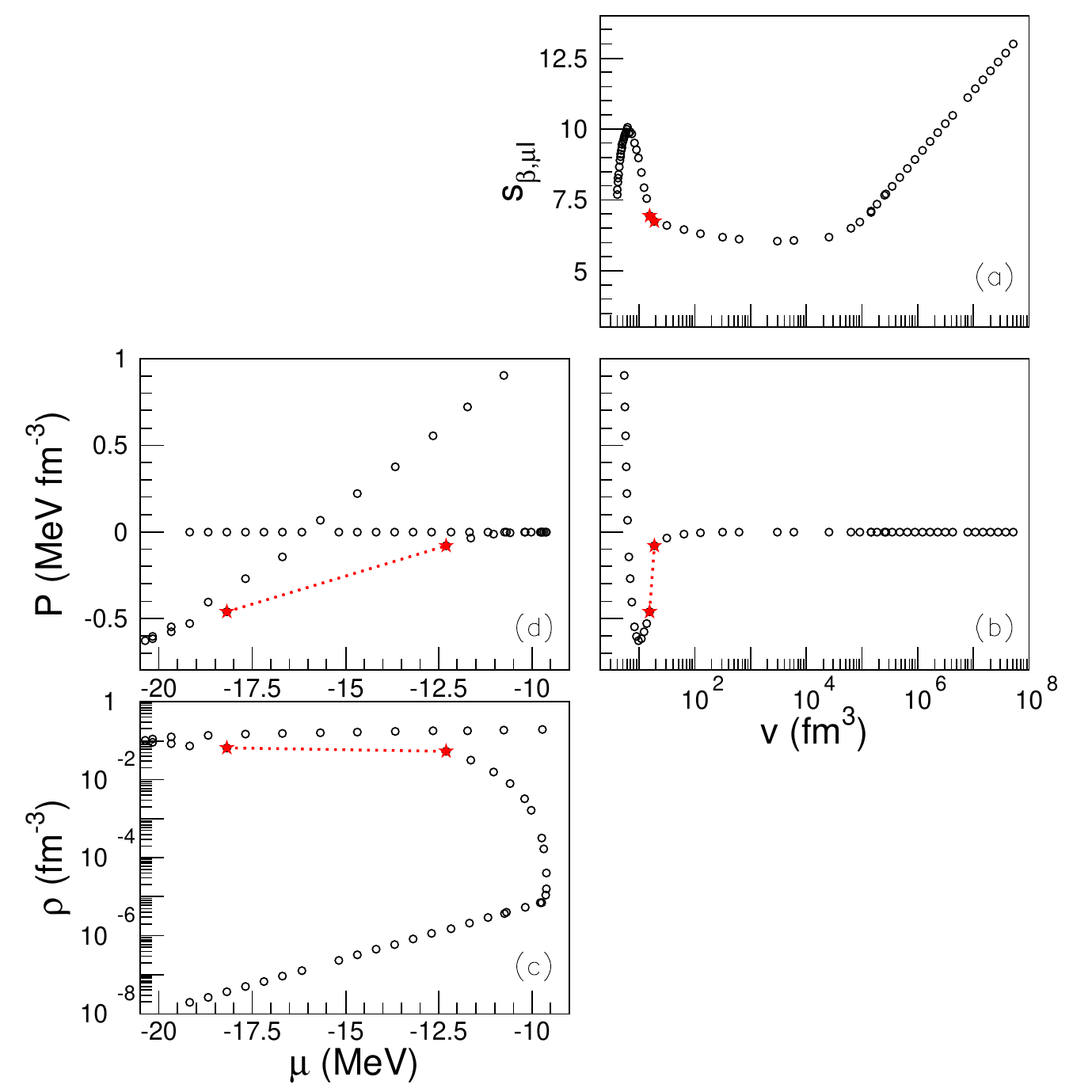}
\caption{ Constrained entropy (a), pressure [(b) and (d)] and chemical
  potential [(c) and (d)] evaluated in the extended NSE model in the
  canonical ensemble at T = 1.6 MeV and asymmetry chemical potential
  $\mu_I$ = 1.68 MeV.  
  Stars and dotted line signal pressure and chemical potential 
  discontinuity.
  Figure taken from ref.~\cite{ensineq}.
 }
\label{fig8}
\end{figure}

The thermodynamics of the extended NSE model is presented in Figure
\ref{fig8}. We can see that the equation of state does not present any
plateau as it would have been the case for a first-order phase
transition.  More surprising, the entropy presents a convex intruder,
the behavior of the equation of state is not monotonic and a clear
back-bending is observed, qualitatively similar to the phenomenon
observed in phase transitions in finite systems~\cite{gross}.  It is
interesting to remark that similar behaviors, with non-monotonic
equations of state and discontinuities in the intensive variables, are
systematically observed in phase transitions with long-range
interactions~\cite{campa}. A consequence of the back-bending in the
equation of state is that this unusual thermodynamics can only be
observed in the canonical ensemble where the density is
controlled. Indeed if the baryon chemical potential was controlled as
it is the case in standard NSE, in the region of the back-bending the
multiple evaluation of the chemical potential would lead to keep only
the solution of minimal free energy. This means that the whole
back-bending region would be jumped over and one would observe a
density discontinuity, that is a first-order phase transition. This
in-equivalence of statistical ensembles is a characteristic feature of
phase transitions with long range interactions. Different model
applications have indeed shown fingerprints of ensemble 
in-equivalence~\cite{campa}, but phenomenological applications are scarce. The NSE
calculation of Figure \ref{fig8} shows that the inhomogeneus baryonic
matter which is produced in the explosion of core-collapse supernova
and in neutron stars is an example of a physical system which displays
this in-equivalence.

\begin{figure}
\includegraphics[width=0.45\columnwidth,height=2.5in,clip]{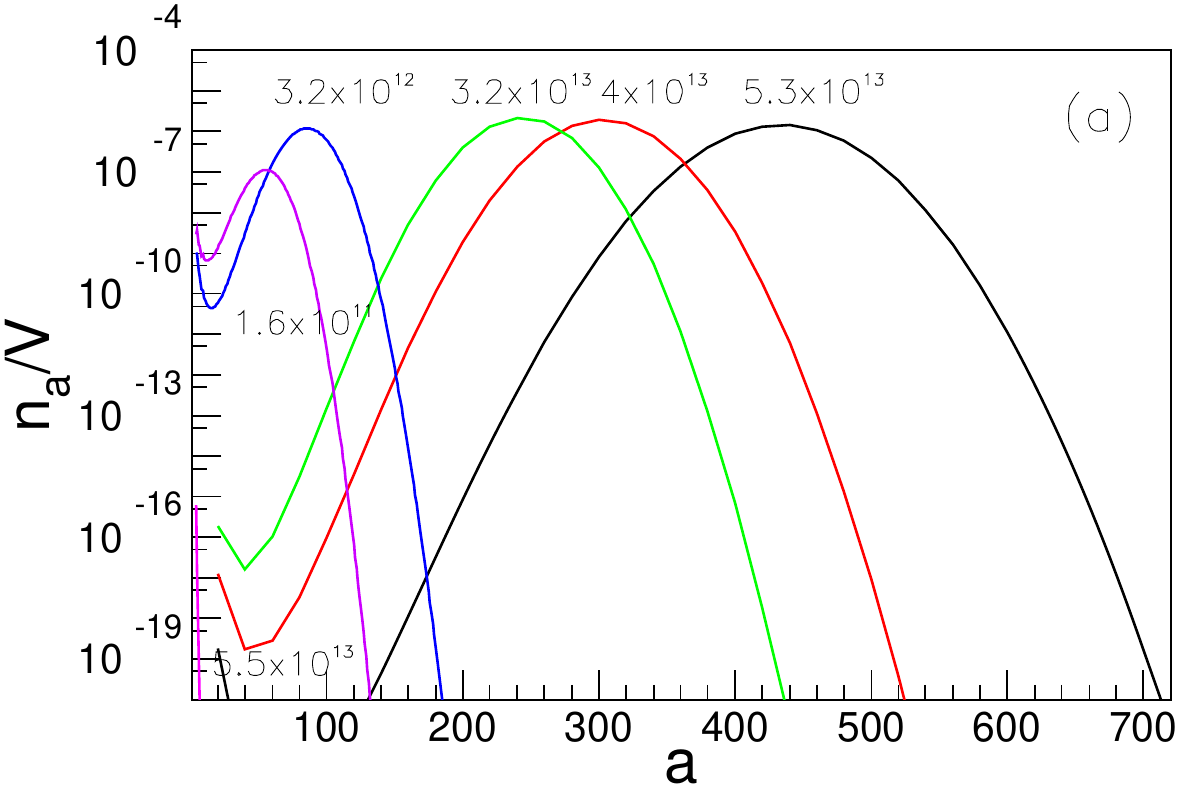}
\includegraphics[width=0.45\columnwidth,height=2.5in,clip]{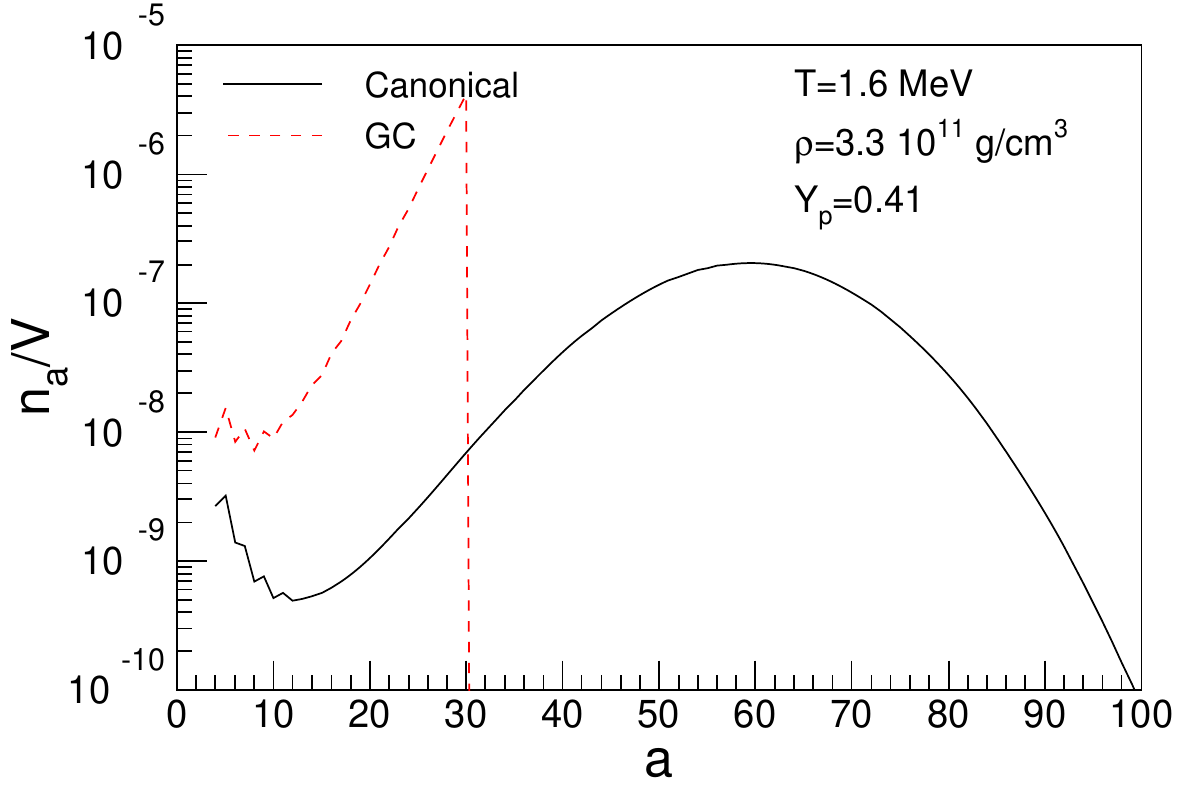}
\caption{ Left side: Cluster distributions as a function of the density 
(expressed in $g/cm^3$) of the extended NSE model in the ensemble 
in-equivalence region in the same thermodynamic conditions than in 
Figure \ref{fig8}. 
Right side: Comparison between canonical (full line) and
grand-canonical (dashed line) predictions for the cluster distribution
in a specific thermodynamic condition relevant for supernova
dynamics. Figure  taken from ref.~\cite{ensineq}. 
 }
\label{fig9}
\end{figure}
From the physical viewpoint, the correct modelization is the canonical one. 
Indeed  if the grandcanonical phase coexistence solution was the preferred 
response at equilibrium inside the in-equivalence region, such solution would 
have been found in a canonical calculation where the coexistence density is 
imposed. On the contrary, once the density is fixed, a dis-homogeneous
 clusterized solution is found in agreement with the expected phenomenology
 associated to the Coulomb frustration.
This is demonstrated in the left part of Figure \ref{fig9}, which
shows the cluster distribution in the in-equivalence region. As a
function of the density, the average cluster size is continuously
changing and can never be assimilated to a portion of the fluid
phase. The right part of the same figure compares the grandcanonical
and canonical cluster distribution in a thermodynamic situation
relevant for supernova physics. We can see that accounting for the
correct thermodynamics has important consequences on the matter
composition. Since the cluster abundances determine the electron
capture rate, which in turn is a capital ingredient for the size of the
homologous core and the cooling process, it is clear that a control of
the phase diagram is important for supernova physics. It is also
interesting to remark that the most probable abundances concern
not so neutron-rich nuclei which are potentially accessible 
 in laboratory experiments: a detailed knowledge of the mass, level density
 and electron capture cross section for these nuclei is thus needed to have
 a reliable description of the equation of state. 
\subsection{In-medium effects}
%
In the extended NSE model presented in the previous section clusters are
 treated as an ideal gas characterized by an empirical energy functional which 
is modified with respect to nuclei in the vacuum only through the excluded volume
 mechanism. This is an improvement compared to standard 
NSE~\cite{NSE,mishustin,blinnikov,souza} in which the Coulomb screening by 
the electron background is the only interaction accounted for, but still 
represents a crude approximation of the effect of the medium~\cite{hempel}. 
Microscopic modifications of the cluster binding energies due to Pauli 
blocking shifts have been computed~\cite{typel} and shown to correctly describe
 the Mott dissolution of deuterons in dense matter, but this heavy numerical
 approach cannot be extended to massive clusters.  
In the original Lattimer-Swesty model~\cite{LS91} the assumption is made that
 the effect of the medium consists in a modification of the cluster surface
 tension, the bulk energy being unaffected. However some interpretations of
 multifragmentation data suggest that bulk terms might also be 
affected~\cite{botvina}.

\begin{figure}
\includegraphics[width=0.9\columnwidth,height=5.in,clip]{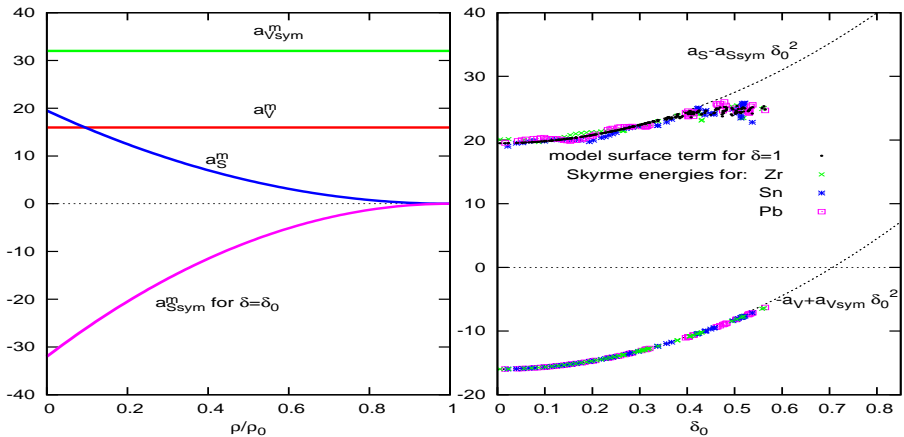}
\vskip -2.5in
\caption{ Left part:  isoscalar and isovector bulk and isoscalar surface term of the
 in-medium modified mass formula extracted from Hartree-Fock calculations
 in the Wigner-Seitz cell. Right part: comparison between the microscopic
 Hartree-Fock volume energy per particle and surface energy per $A^{2/3}$ and the parameterization eq.(\ref{pana}) for three
 isotopic chains as a function of the bulk proton-neutron asymmetry. 
The model surface term is given by 
$\left [ a_{S}^m(\rho) -a_{Ssym}^m(\rho,\delta){\delta_0}^2\right ]$, 
see eq.~(\ref{pana}). 
 }
\label{fig4}
\end{figure}

To microscopically account for in-medium modifications of the cluster
 binding energies, we have realized a systematic set of Hartree Fock 
calculations in the Wigner Seitz cell for various isotopic chains 
with a total neutron number ranging from the stability line up 
to $N\approx 3000$. The resulting microscopic nuclear energy, consisting of a volume and a surface term, is parameterized with 
an in-medium modified mass formula
%
%
%
%
%
\begin{equation}
E_{HF}=\epsilon^{HF}(\rho_n,\rho_p) (V_{WS}-V_{cl}) -\left [ a_V^m(\rho) -
a_{Vsym}^m(\rho,\delta)\delta_0^2\right ] A+
\left [ a_{S}^m(\rho) -a_{Ssym}^m(\rho,\delta){\delta_0}^2\right ] A^{2/3} \label{pana}
\end{equation}
%
where $V_{WS}$ is the volume of the calculation cell and $\rho_n,\rho_p,\rho,\delta$ is the neutron, proton, total density and asymmetry $\delta=(\rho_n-\rho_p)/\rho$ of the free nucleons constituting the dripping gas. The cluster mass and charge $A$, $Z$ 
and the gas densities 
are extracted through a Wood-Saxon fit of the local density. 
The bulk density $\rho_0$ and asymmetry $\delta_0=(\rho_{0n}-\rho_{0p})/\rho_0$ in the model are approximated by analytical functions of $A,Z$, while $V_{cl}=A/\rho_0$. 
The resulting in-medium parameters obtained from a preliminary fit of the microscopic Hartree-Fock results through eq.(\ref{pana}) 
are shown in the left part of Figure \ref{fig4}, while the quality of the fit is displayed in the right part of the same figure. 

To obtain the fit shown in Fig.\ref{fig4} we have employed the ansatz 
\begin{equation}
a_S^m(\rho ) = a_S \left(1-\frac{\rho}{\rho_0}\right)^{\gamma_1} \; \; ;\; \;
 a_{Ssym}^m(\rho,\delta ) = a_{Ssym} \left (1-\frac{\rho\delta}{\rho_0\delta_0}\right )^{\gamma_2}
\end{equation}
with $\gamma_1=\gamma_2 =2$  and  with $ a_{Ssym}=-32$ MeV.
A dependence of $a_{Ssym}^m$ on $\rho_0$,$\delta_0$ is implicit. 
We can see that the in-medium modified liquid drop expression eq.(\ref{pana}) 
provides a reasonably accurate description of the energy modification due to
 the external gas. In agreement with the conjecture  of ref.~\cite{LS91}, the
 bulk terms are completely unaffected by the external medium ($a_V^m(\rho)=a_V$=cst., $a_{Vsym}^m(\rho,\delta)=a_{Vsym}$=cst.), 
while important effects on
 the surface properties of nuclei are seen. 
The implementation of such an improved energy functional in the extended
 NSE model is in progress.

\section{Supra-saturation densities and the strangeness phase transition}

If it is well admitted that hyperonic and deconfined quark matter could exist
in the inner core of neutron stars, a complete understanding of the
composition and equation of state of dense matter is far from being achieved.
Concerning hyperons, simple energetic considerations suggest that they should
be present at high density~\cite{Glendenning82}.  However, in the standard
picture the opening of hyperon degrees of freedom leads to a considerable
softening of the equation of state~\cite{Glendenning82,Baldo99, Vidana00,
  Djapo10,Schulze11,Massot12}, which in turns leads to
maximum neutron star masses smaller than the observed masses of many
neutron stars, in particular the high values obtained in recent
observations~\cite{demorest}. This puzzling situation implies that the
hyperon-hyperon and hyperon-nucleon couplings must be much more repulsive at
high density than presently assumed~\cite{Hofmann00,Bonanno11,Weissenborn11,
Bednarek11, Lastowiecki11, Oertel12}, and/or that something
is missing in the present modelization.

\begin{figure}
\begin{center}
\includegraphics[angle=0, width=0.45\columnwidth,height=5.in]{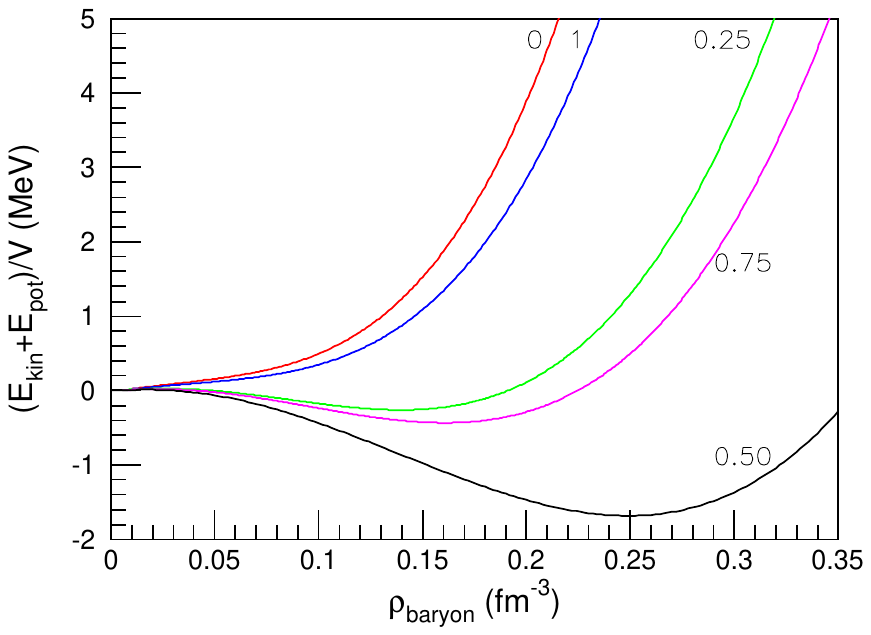}
\includegraphics[angle=0, width=0.45\columnwidth,height=4.in]{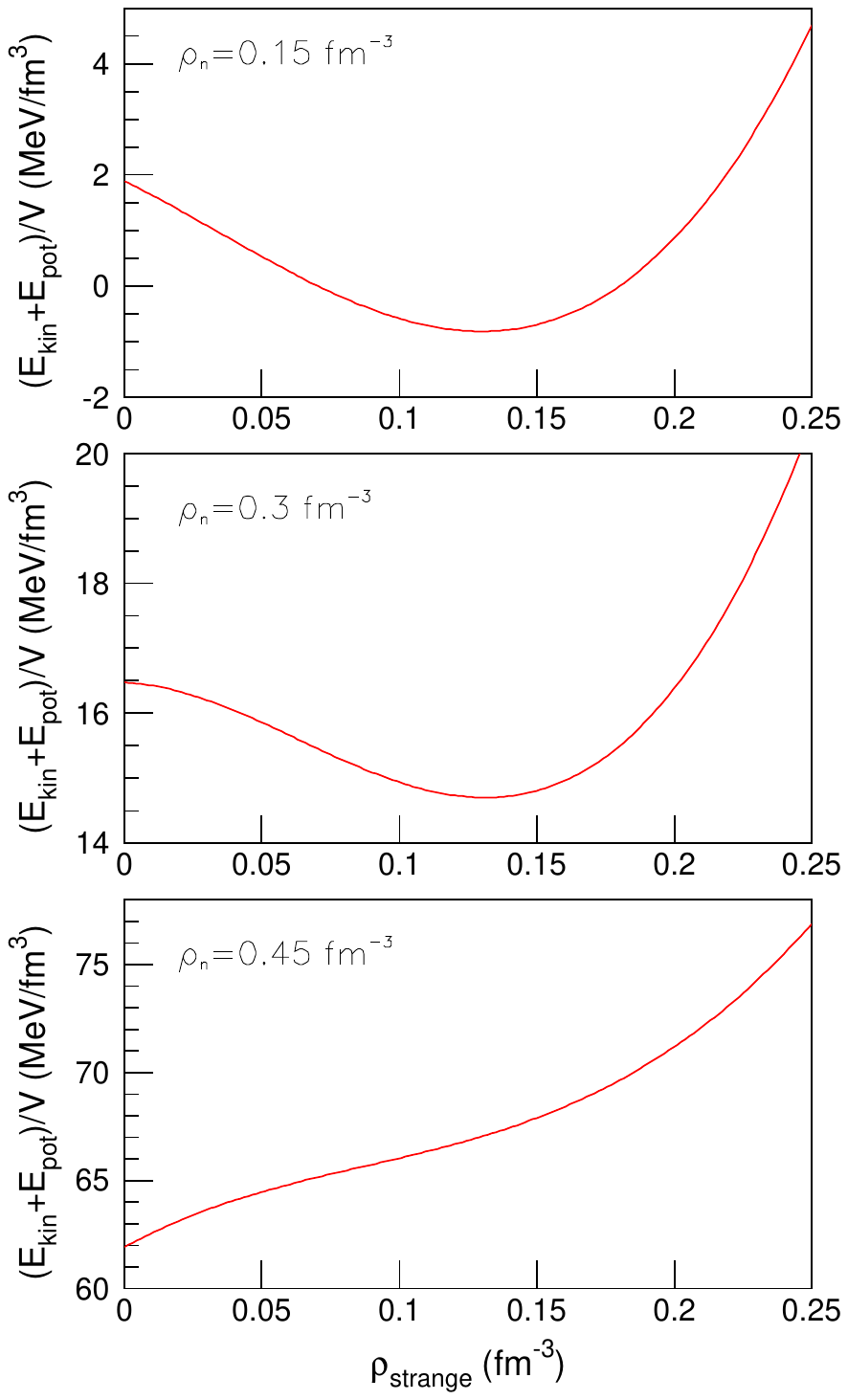}
\end{center}
\caption{Left part: energy density as a function of the total baryon
  density for a $(n,\Lambda)$ mixture at $T=0$ with different
  $\Lambda$ fractions with the energy functional proposed in 
ref.~\cite{Balberg97}. Right part: same as a function 
of the strange density at fixed neutron densities.} 
\label{fig:energy_vs_rhob}
\end{figure}

The generic presence of attractive and
repulsive couplings suggests the existence, in a model-independent manner, of a
phase transition involving strangeness. Let us consider a simplified 
situation where matter is uniquely composed of neutrons and $\Lambda$-hyperons. 
The energy density $\epsilon=\epsilon_{kin}+\epsilon_{pot}$ at
zero temperature obtained with  the
energy density functional proposed by Balberg and Gal~\cite{Balberg97}
\begin{eqnarray}
\epsilon_{pot}\left(\rho_n,\rho_\Lambda\right)&=& 
\frac{1}{2}\left [\left(a_{NN}+b_{NN}\right)\rho_n^2+c_{NN}\rho_n^{\delta+1}\right]+
\frac{1}{2}\left [a_{\Lambda\Lambda}\rho_\Lambda^2
+c_{\Lambda\Lambda}\rho_\Lambda^{\gamma+1}\right]\nonumber \\
&+&a_{\Lambda n} \rho_n \rho_{\Lambda} + c_{\Lambda n}\frac{\rho_n \rho_\Lambda}{\rho_n+\rho_\Lambda}
\left [ \rho_n^\gamma+\rho_\Lambda^\gamma\right ], \label{balb}
\end{eqnarray}
 for different
hyperon fractions $Y_{\Lambda}=\rho_{\Lambda}/(\rho_{\Lambda}+\rho_n)$
is represented in the left part of  Fig.~\ref{fig:energy_vs_rhob}. For
all numerical applications, the parameter values from parameterization
BG I, see ref.~\cite{lambda_noi} are used.  
We can see that, due to the attractive part of the $n\Lambda$
coupling, a mixture of the two particle species admits a bound state at
finite density.  This rather general feature of
the energy density is found within other parameterizations and other
models, e.g. the parameterization of the energy density functional from
G-matrix calculations by Vidana et al.~\cite{Vidana10}. 
The existence of a minimum in the energy functional for symmetric matter
means that such a state represents the stable matter phase at zero temperature.
On the other side, the vanishing density gas phase is always the stable phase
in the limit of infinite temperature. 
This implies that a dilute-to-dense phase change
in strange matter has to be expected on very general grounds.
The right part of Fig.\ref{fig:energy_vs_rhob} shows that the minimum, 
for a certain interval of neutron densities, is also observed as a function 
of the $\Lambda$ density. This is again due to the interplay between the 
attractive
$\Lambda-\Lambda$ interaction required by hypernuclear data, and the
repulsion at high density required by the two-solar mass neutron star
observation. This information means that the expected phase change is
expected to have strangeness contributing to the order parameter.  

The presence of a minimum is however not sufficient to discriminate 
between a smooth cross-over and a phase transition. 
In the next section we will therefore demonstrate the existence of a 
phase transition by explicitly calculating the $n\Lambda$ phase diagram.
\subsection{Neutron-$\Lambda$ phase diagram}
 First-order transitions
are signaled by an instability or concavity anomaly in the mean-field
thermodynamic potential, which has to be cured by means of the Gibbs phase
equilibrium construction at the thermodynamic limit.   At
zero temperature, one thermodynamic
potential is given by the total energy
\begin{equation}
\epsilon_{tot}\left (\rho_n,\rho_\Lambda \right)=\epsilon_{pot}
\left (\rho_n,\rho_\Lambda \right)+\epsilon_{kin}
\left (\rho_n,\rho_\Lambda \right)+\left (\rho_n m_n +
\rho_\Lambda m_\Lambda\right ) c^2 
\end{equation}
The spinodal region is then
recognized as the locus of negative curvature of the energy surface,
$C_{min}<0$. The corresponding eigenvector defines a direction in the density
space given by
\begin{equation}
\frac{\rho_n}{\rho_{\Lambda}}=\frac{C_{n \Lambda}}{C_{min}-C_{nn}}=
\frac{C_{min}-C_{\Lambda \Lambda}}{C_{\Lambda n}}
\end{equation}
with $C_{ij}=\partial^2 e_{tot}/\partial \rho_i \partial \rho_j$.
This instability direction physically represents the chemical composition of
density fluctuations which are spontaneously and exponentially amplified in
the unstable region in order to achieve phase separation, and gives the order
parameter of the associated phase transition.

\begin{figure}
\begin{center}
\includegraphics[angle=0, width=0.4\columnwidth]{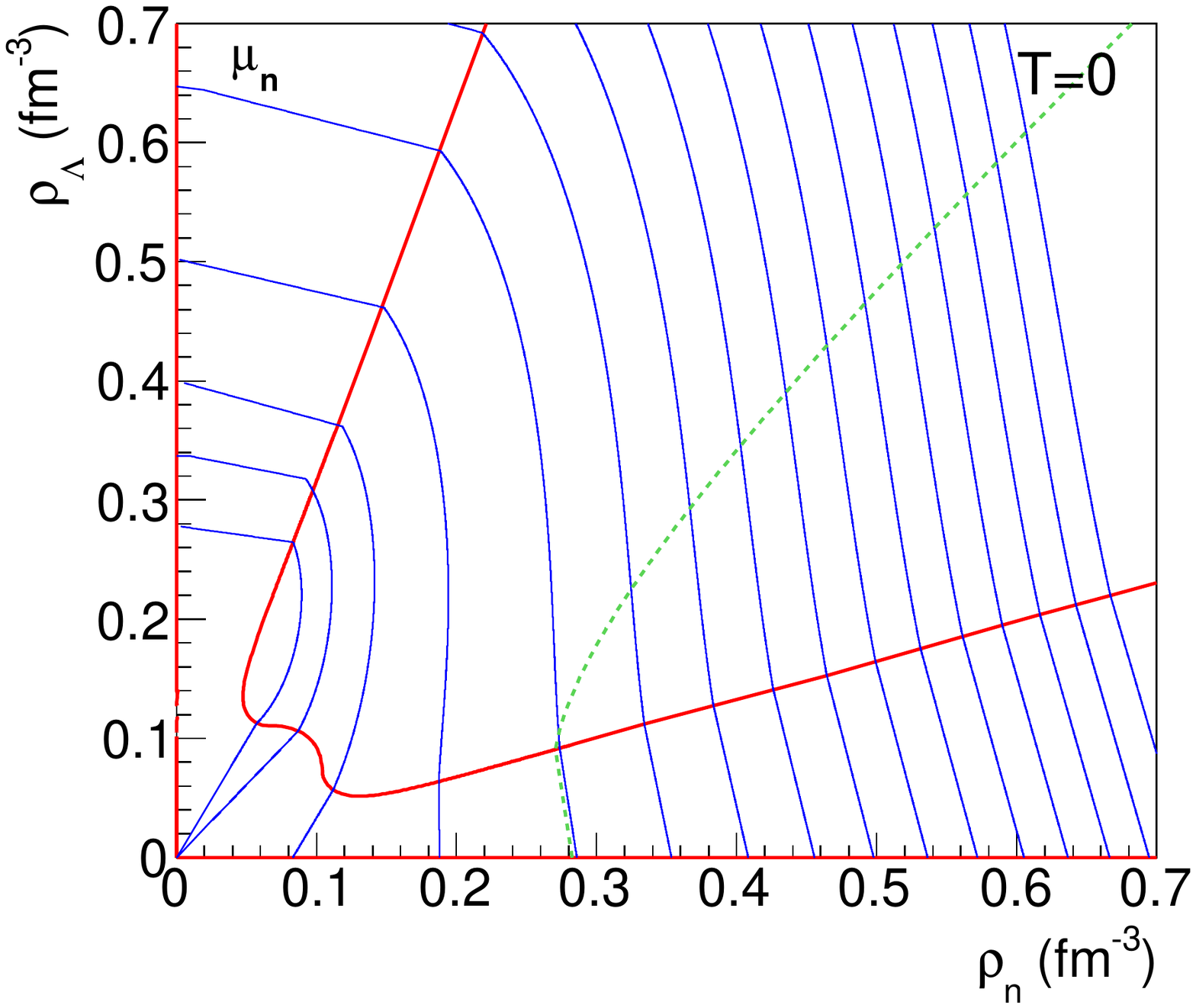}
\includegraphics[angle=0, width=0.4\columnwidth]{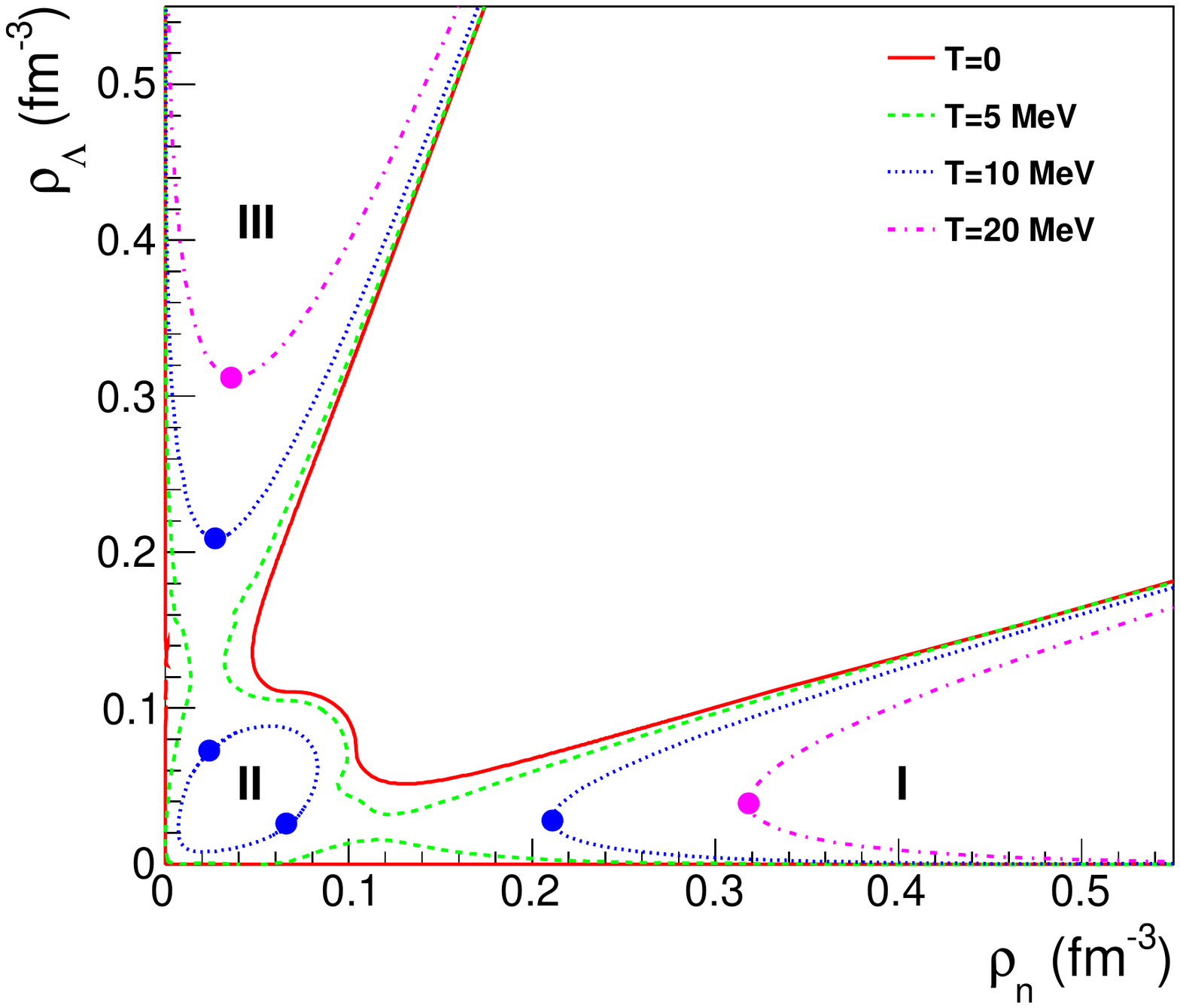}
\end{center}
\caption{Left part: $(n,\Lambda)$ mixture at $T=0$:
Borders of the phase coexistence region (red thick lines),
constant-$\mu_n$ paths after Maxwell construction (blue thin lines)
and $\mu_n=\mu_{\Lambda}$ trajectory after Gibbs construction 
(dark green dotted line).  Right part:
  borders of the phase coexistence region of the 
$n\Lambda$ mixture for different temperatures, $T$=0, 5, 10, 20 MeV. 
The full circles mark the critical points. Figures taken 
from ref.~\cite{lambda_noi}.
}
\label{fig:coex_contoursmun}
\end{figure}

The zero temperature instability region of the $n\Lambda$ mixture is shown in
the left part of Fig.~\ref{fig:coex_contoursmun}.
We can see that a large portion of the phase diagram is concerned by the
instability. We can qualitatively distinguish three regions characterized by
different order parameters. Below nuclear saturation density, we observe an
isoscalar $\rho_n\approx\rho_\Lambda$ instability, very close to ordinary
nuclear liquid-gas, with $\Lambda$'s playing the role of protons.  
Increasing neutron density the phase separation progressively changes towards
the strangeness $\rho_S=-\rho_\Lambda$ direction: the two stable phases
connected by the instability have close baryon densities but a very different
fraction of $\Lambda$'s. For high $\rho_{\Lambda}$ we observe the same
behavior with the roles of neutrons and $\Lambda$-hyperons exchanged.  The
part of the phase diagram at high neutron density comprises the region
physically explored by supernova and neutron star matter, which is
characterized by chemical equilibrium for reactions implying strangeness,
$\mu_S=\mu_n-\mu_\Lambda=0$. The strangeness equilibrium trajectory is
represented by a dashed line in Fig.\ref{fig:coex_contoursmun}.  This physically
corresponds to the sudden opening of strangeness, observed in many
modelizations of neutron star matter (see e.g.~\cite{Balberg97, Massot12}).

Let us stress that  the existence of a strangeness phase transition at 
high density  is not a general model-independent feature, although many 
models show it. There are others, for instance the
G-matrix models of ref.~\cite{Vidana10}, 
which do not show an instability in
this region.   The reason is essentially due to the absence of 
 a $\Lambda\Lambda$ interaction in these
microscopic calculations. Owing to the lack of reliable information on
the hyperonic interactions, which would discriminate between different models,
we cannot affirm the existence of the strangeness phase transition, related to
this instability, but, turning the argument around, the presence of this phase
transition in a physical system would allow to learn much about the shape of
the interaction.

\begin{figure}
\begin{center}
\includegraphics[angle=0, width=0.45\columnwidth]{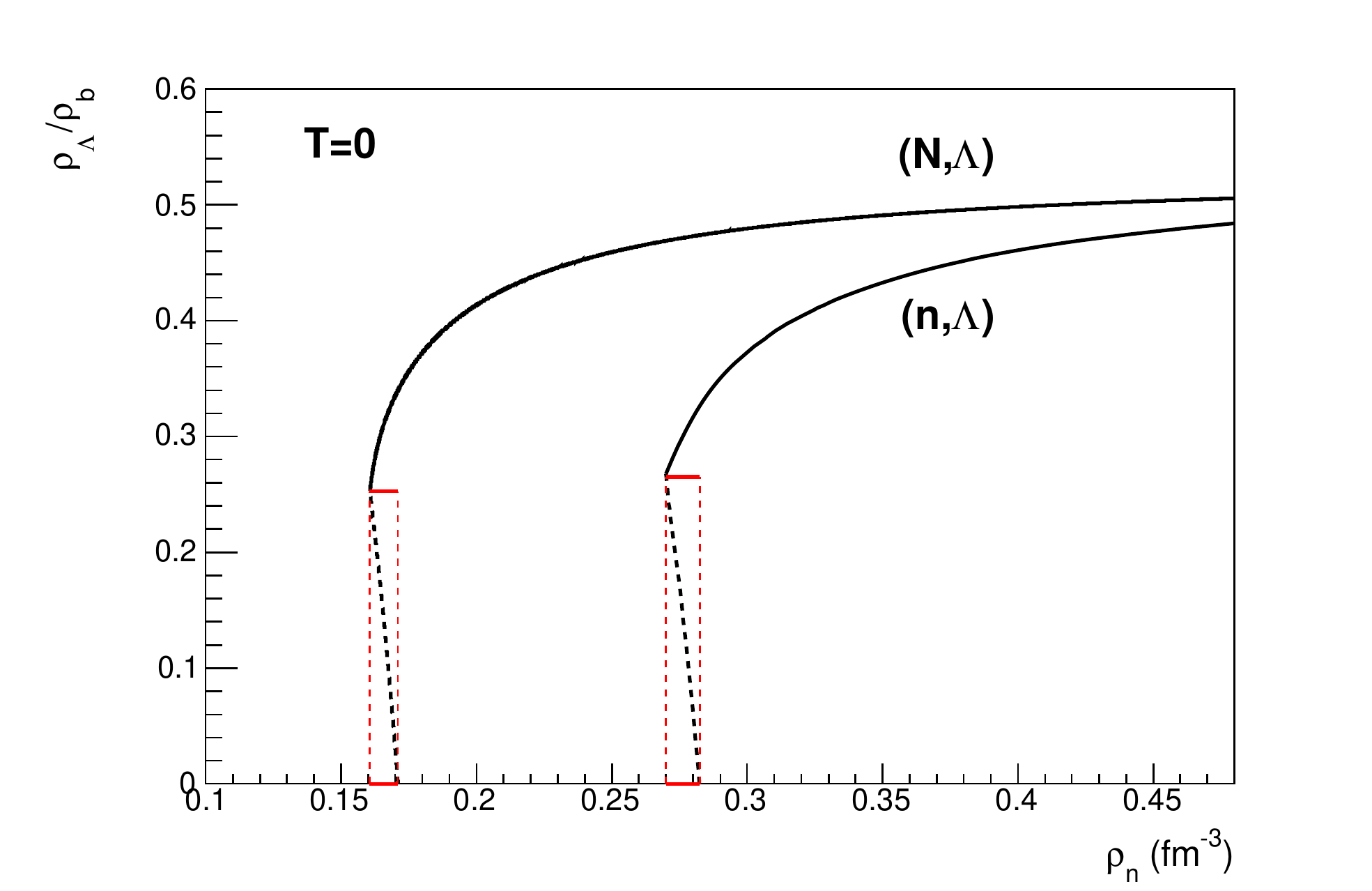}
\includegraphics[angle=0, width=0.45\columnwidth,height=5.in]{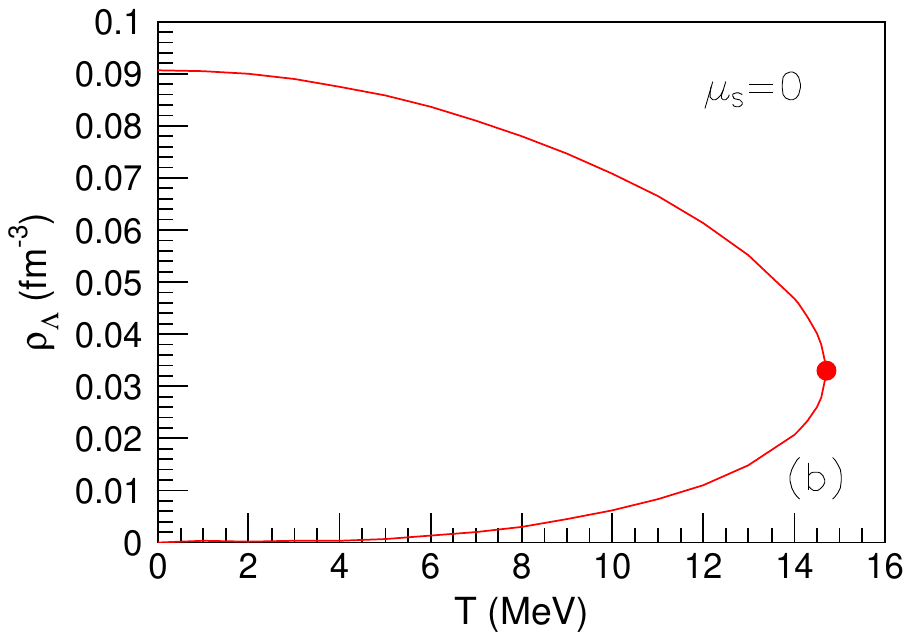}
\end{center}
\caption{ 
  Left: $\Lambda$ fraction as a function of the baryon density within 
  the condition $\mu_n=\mu_\Lambda$ for a 
  $n\Lambda$ and, respectively, a symmetric $\rho_n=\rho_p$ 
  $n p \Lambda$ mixture 
  at $T$=0.
  The full lines correspond to the stable mean-field results; 
  the dashed lines illustrate the Gibbs constructions; 
  the horizontal lines indicate the relative amount of $\Lambda$-hyperons 
  at the frontiers 
  of the phase-coexistence domain.
  Right: $\Lambda$ density for the two coexisting phases as a function of 
  temperature for $\mu_S=0$. 
  Figures taken from ref.~\cite{lambda_noi}.
}
\label{fig:lambdas}
\end{figure}

Up to now we have only presented results at zero temperature. The extension to
finite temperature, as it is needed for supernova matter, is relatively simple
in the mean-field approximation.  The appropriate thermodynamic potential at
non-zero temperature is given by the Helmholtz free energy
\begin{equation}
f_T\left ( \rho_n,\rho_\Lambda\right)= \epsilon_{tot}
\left ( \rho_n,\rho_\Lambda\right)-Ts\left ( \rho_n,\rho_\Lambda\right)
\end{equation}
where $s$ is the mean-field entropy density.  Chemical potentials can be
obtained by differentiating this expression.
   Then the Gibbs construction is performed from the combined analysis of 
$\mu_n(\rho_n)$ at constant
$\mu_\Lambda$, and $\mu_\Lambda(\rho_\Lambda)$ at constant $\mu_n$.

The phase diagram as a function of temperature is presented in
the right panel of Fig.~\ref{fig:coex_contoursmun}.  
This phase diagram exhibits different interesting
features.  We can see that the three regions that we have tentatively defined
at zero temperature appear as distinct phase transitions at finite
temperature. The first phase transition (zone II in 
Fig.\ref{fig:coex_contoursmun} - right panel)
separates a low density gas phase from a high density more symmetric liquid
phase, very similar to ordinary liquid-gas.  The second one (zone III in
Fig.\ref{fig:coex_contoursmun} - right panel) 
reflects the instability of dense strange matter towards
the appearance of neutrons and has an almost symmetric counterpart (zone I
in Fig.\ref{fig:coex_contoursmun} - right panel) 
in the instability of dense neutron matter towards
the formation of $\Lambda$-hyperons. 
Up to a certain temperature, this latter phase transition is explored
by the $\mu_n=\mu_\Lambda$ trajectory, meaning that it is expected to occur in
neutron stars and supernova matter. At variance with other known phase
transitions in nuclear matter, this transition exists at any temperature and
is not limited in density; it is always associated 
with a critical point, which moves towards high density as the temperature
increases. This means that criticality should be observed in hot supernova
matter, at a temperature which is estimated as $T_c=14.8$ MeV in the present
schematic model.

The consequences of these findings for the composition of neutron star matter
are drawn in Fig.~\ref{fig:lambdas}. The left panel shows the $\Lambda$ fraction
$Y_\Lambda=\rho_\Lambda/(\rho_n+\rho_\Lambda)$ as a function of the baryon
density under the condition $\mu_n=\mu_\Lambda$. The crossing of the
mixed-phase region with increasing neutron density implies that, as soon as
the lower density transition border is crossed, the system has to be viewed as
a dis-homogeneous mixture of macroscopic regions composed essentially of
neutrons, with other macroscopic regions with around $25\%$ of hyperons. The
extension of the $\Lambda$-rich zone increases with density until the system
exits the coexistence zone, and becomes homogeneous again.
The density values associated to the phase transition slightly change with the
energy functional assumed. To give a simple example, we present on the same
Fig.\ref{fig:lambdas} a calculation where $\Lambda$'s are put in equilibrium
with symmetric $\rho_n=\rho_p$ non-strange matter, that is eq.(\ref{balb})
 is still used, but the symmetry energy in the
nucleon sector $b_{NN}$ is put to zero~\cite{Balberg97}.  We can see that only
small differences are observed in the strange phase transition considering
pure neutron matter or symmetric matter.  

On the right panel of Fig.~\ref{fig:lambdas} the $\Lambda$-densities for the
two coexisting phases are displayed as a function of temperature, again under
the condition of $\mu_n = \mu_\Lambda$, physically relevant for neutron star
and supernova matter. The extension of the coexistence region decreases with
increasing temperature. The latter finally disappears at $T_c = 14.8$
MeV. This well illustrates the fact  that the critical point of the
strangeness phase transition moves to higher density, crossing the physical
line $\mu_n = \mu_\Lambda$ at a given temperature.

\section{Conclusions}
In this contribution we have analyzed the phase diagram of stellar matter 
as it is produced in core-collapse supernovae and proto-neutron stars in 
the thermodynamical conditions where such matter can be described by 
nucleonic and hadronic degrees of freedom.

At baryonic densities lower than nuclear saturation, we have shown on 
very general arguments that the first-order phase transition of neutral 
nuclear matter is quenched in stellar matter due to the effect of Coulomb 
frustration.
As a consequence, matter is expected to be clusterized even at finite 
temperature in the whole sub-saturation regime.

The composition and thermodynamics of such clusterized matter has been 
explicitly worked out in the framework of a quasi-particle model where 
inter-particle interactions are modeled through excluded volume. 
Coulomb correlations with the neutralizing electron gas are accounted 
for in the Wigner-Seitz approximation and residual nuclear correlations 
are included in an effective way through an in-medium modification of 
the cluster energy functional with parameters fitted from Hartree-Fock 
calculations with a 
realistic effective interaction. 
The same formalism 
is used to evaluate the free nucleon self-energies.

The expectation of the phase transition quenching is confirmed by this
model, which shows that the single nucleus approximation widely used
is a poor approximation at finite temperature and a large distribution
of cluster species has to be included in supernova modeling. In
addition to that, the Coulomb frustration effect is seen to
drastically modify the functional relation between pressure, chemical
potential and density with respect with a first-order phase
transition. Specifically, backbendings are observed in the equations
of state which cannot be correctly evaluated in the grandcanonical 
ensemble but require a canonical evaluation of the partition sum.

In the supersaturation regime, we have used the same mean field 
approximation with effective phenomenological energy functionals in order 
to describe the characteristics of the phase diagram in the presence of 
strangeness. 

  We have shown
that already a simple system composed of neutrons and $\Lambda$ hyperons 
in thermal equilibrium presents a complex phase diagram with first and 
second order phase transitions. Some of these phase transitions are probably
 never
explored in physical systems. However, a possible phase transition at
super-saturation baryon densities, from non-strange to strange matter is
expected to be observed both in the inner core of neutron star and in the
dense regions of core-collapse supernova. For this latter phenomenology, a
critical point is predicted and the associated critical opalescence could have
an impact on supernova dynamics~\cite{jmargue}.    
 The immediate consequence of that is
that the opening of hyperon channels at high density should not be viewed as a
continuous (though abrupt) increase of strangeness in the matter, observed in
many models of hyperonic matter, cf e.g.~\cite{Massot12}, but rather as the
coexistence of hyperon-poor and hyperon-rich macroscopic domains.

An extension of  this simple model to include all possible hyperons and 
resonances is in progress.

\end{document}